\documentclass[aps,prb,reprint,superscriptaddress,longbibliography,showpacs,floatfix]{revtex4-1}

\usepackage{graphicx}
\usepackage{epstopdf}
\usepackage{xcolor}
\usepackage{soul}
\usepackage{amsmath}
\usepackage{upgreek}

\begin{document}

%\preprint{}

\title{Magnetic properties and field-driven dynamics of chiral domain walls\\ in epitaxial Pt/Co/Au$_x$Pt$_{1-x}$ trilayers}


\author{Kowsar~Shahbazi}
\affiliation{School of Physics and Astronomy, University of Leeds, Leeds LS2 9JT, United Kingdom}

\author{Ale\v{s}~Hrabec}
\altaffiliation{Present Address: Mesoscopic Systems, Paul Scherrer Institut, ODRA/113, 5232 Villigen, Switzerland.}
\affiliation{School of Physics and Astronomy, University of Leeds, Leeds LS2 9JT, United Kingdom}
\affiliation{Laboratoire de Physique des Solides, CNRS, Universit\'{e} Paris-Sud, Universit\'{e} Paris-Saclay, 91405 Orsay Cedex, France}

\author{Simone~Moretti}
\affiliation{Departamento Fisica Aplicada, University of Salamanca, Plaza de los Caidos s/n E-37008, Salamanca, Spain}
\affiliation{Fachbereich Physik, Universit\"{a}t Konstanz, 78457 Konstanz, Germany}

\author{Michael~B.~Ward}
\affiliation{Leeds Electron Microscopy and Spectroscopy Centre, School of Chemical and Process Engineering, University of Leeds, Leeds LS2 9JT, United Kingdom}

\author{Thomas~A.~Moore}
\affiliation{School of Physics and Astronomy, University of Leeds, Leeds LS2 9JT, United Kingdom}

\author{Vincent~Jeudy}
\affiliation{Laboratoire de Physique des Solides, CNRS, Universit\'{e} Paris-Sud, Universit\'{e} Paris-Saclay, 91405 Orsay Cedex, France}

\author{Eduardo~Martinez}
\affiliation{Departamento Fisica Aplicada, University of Salamanca, Plaza de los Caidos s/n E-37008, Salamanca, Spain}

\author{Christopher~H.~Marrows}\email[Email:~]{c.h.marrows@leeds.ac.uk}
\affiliation{School of Physics and Astronomy, University of Leeds, Leeds LS2 9JT, United Kingdom}

\date{\today}

\begin{abstract}
Chiral domain walls in ultrathin perpendicularly magnetised layers have a N\'{e}el structure stabilised by a Dzyaloshinskii-Moriya interaction (DMI) that is generated at the interface between the ferromagnet and a heavy metal. Different interface materials or properties are required above and below a ferromagnetic film in order to generate the structural inversion asymmetry needed to ensure that the DMI arising at the two interfaces does not cancel. Here we report on the magnetic properties of epitaxial Pt/Co/Au$_x$Pt$_{1-x}$ trilayers grown by sputtering onto sapphire substrates with 0.6~nm thick Co. As $x$ rises from 0 to 1 a structural inversion asymmetry is progressively generated. We characterise the epilayer structure with x-ray diffraction and cross-sectional transmission electron microscopy, revealing (111) stacking. The saturation magnetization falls as the proximity magnetisation in Pt is reduced, whilst the perpendicular magnetic anisotropy $K_\mathrm{u}$ rises. The micromagnetic DMI strength $D$ was determined using the bubble expansion technique and also rises from a negligible value when $x=0$ to $\sim 1$~mJ/m$^2$ for $x = 1$. The depinning field at which field-driven domain wall motion crosses from the creep to the depinning regime rises from $\sim 40$ to $\sim 70$~mT, attributed to greater spatial fluctuations of the domain wall energy with increasing Au concentration. Meanwhile, the increase in DMI causes the Walker field to rise from $\sim 10$ to $\sim 280$~mT, meaning that only in the $x = 1$ sample is the steady flow regime accessible. The full dependence of domain wall velocity on driving field bears little resemblance to the prediction of a simple one-dimensional model, but can be described very well using micromagnetic simulations with a realistic model of disorder. These reveal a rise in Gilbert damping as $x$ increases.
\end{abstract}


\maketitle

\section{Introduction}

Research into magnetic thin films has made huge progress over the last few decades, largely due to the additional functionality emerging at interfaces between a magnetic film of few or sub-nanometer thickness and its surroundings. In addition to a wide range of novel magnetotransport effects \cite{Hoffmann2015}, the magnetic properties are profoundly affected by the choice of materials with which those interfaces are made and the nature and quality of the interfaces. Emergent properties span both the statics and dynamics of the magnetization. One of the first to be discovered was the perpendicular magnetic anisotropy (PMA) \cite{BlandHeinrich1PMA}, which can be strong enough to cause a spin-reorientation transition and lift the magnetisation out of the film plane in the ground state. Forming an interface with a metal close to satisfying the Stoner criterion can lead to proximity magnetism \cite{Antel1999}, which can account for a significant proportion of the total moment if the magnetic layer is very thin. Long-predicted \cite{fert1991magnetic}, but only recently observed \cite{chen2013novel}, is the interfacial Dzyaloshinskii-Moriya interaction (DMI), which arises due to inversion symmetry breaking and favours chiral magnetic states \cite{benitez2015magnetic,tetienne2015nature}. Meanwhile, the damping of magnetisation dynamics can be enhanced by spin-pumping into the surrounding layers \cite{Tserkovnyak2002}.

Magnetising a film perpendicular to its plane is a highly demagnetising configuration and domain structures \cite{hubertandschaefer}, separated by domain walls (DWs), commonly occur. Whilst magnetostatics alone favours a Bloch wall configuration, the interfacial DMI takes a form that would prefer a N\'{e}el structure with a fixed chirality \cite{Thiaville_DMI}. This is important for the latest forms of racetrack memory that exploit PMA materials \cite{Parkin2015}, since the DWs are driven by spin-orbit torques \cite{manchon2014spin} that require a component of magnetization collinear to the electron flow. This is not trivial to arrange by other means and it explains, for example, why Bloch walls are insensitive to this type of torque \cite{khvalkovskiy2013matching}, and why the wall chirality determines the direction of wall motion under current \cite{emori2013current,ryu2014chiral}.

Many of these effects, for instance the PMA and DMI, require strong spin-orbit coupling (SOC) and so it is heavy metals that are commonly used to form the interfaces with the ultrathin magnetic layer, usually in polycrystaline multilayers prepared by sputtering. Nevertheless, properties such as the DMI are extremely sensitive to the interface quality \cite{hrabec2014measuring,Wells2017}. To avoid such ambiguities and in order to explore this systematically one has to grow such materials in a controlled way. Here we experimentally investigate epitaxial layers of Pt/Co/Au$_x$Pt$_{1-x}$ prepared by sputtering onto single crystal sapphire substrates \cite{mihai}. Pt is a common pairing with Co that has strong PMA and DMI\cite{hrabec2014measuring}, and is close to satisfying the Stoner criterion\cite{Stoner1938}, giving strong proximity magnetism. The choice of Au, a heavy element with fully filled $5d$ band, is motivated by a negligible proximity effect \cite{wilhelm2004magnetic,ryu2014chiral} and the expectation of a small DMI arising at Co/Au interface \cite{kashid2014dzyaloshinskii,yang2015anatomy}. Gold therefore serves as a textbook element giving the opportunity to study the effect of controlled broken symmetry on either side of the ferromagnet and its importance for the various effects arising at these interfaces.

\section{Epilayer growth and characterization}

\begin{figure}[tb]
  \includegraphics[width=8.5cm]{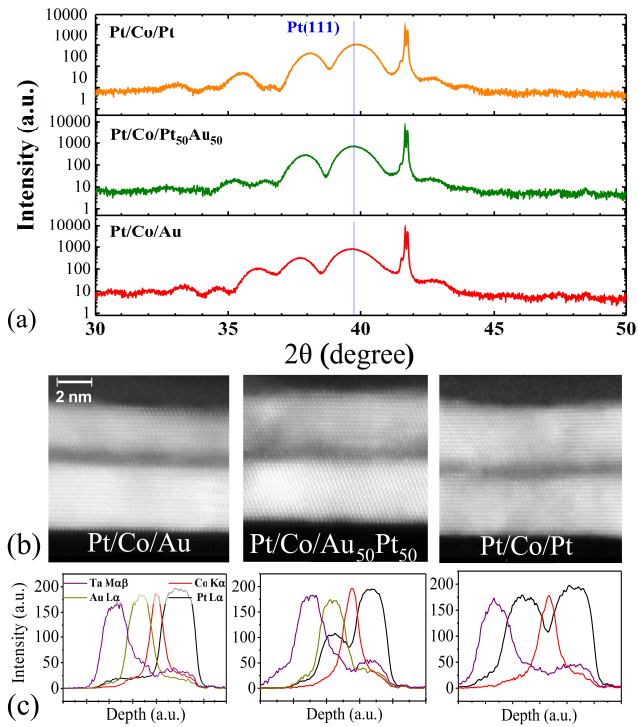}
  \caption{Structural characterization of Pt/Co/Au$_x$Pt$_{1-x}$ trilayers. (a) X-ray diffraction spectra confirming the fcc (111) orientation of the samples. (b) Dark field HRTEM cross-section images. (c) EDX spectra showing distinguishable layers with possible intermixing at the interfaces.
  \label{structure}}
\end{figure}

The trilayers of Pt(3~nm)/Co(0.6~nm)/Au$_x$Pt$_{1-x}$ (3~nm), in which $0 \leq x \leq 1$, were prepared by sputtering at high temperatures, as described in Ref.~\onlinecite{mihai}. The seed Pt layer was sputtered directly on a C-plane sapphire substrate at $500^\circ$C followed by Co at $100^\circ$C. The Au$_x$Pt$_{1-x}$ layer was grown by co-sputtering from Pt and Au targets at $100^\circ$C in a way that the sputtering powers were adjusted to keep the rate $\sim 1$~\AA/s. The multilayers were capped with $\sim$2~nm of Ta, which is polycrystalline.

\begin{figure*}[t!]
  \includegraphics[width=17.8cm]{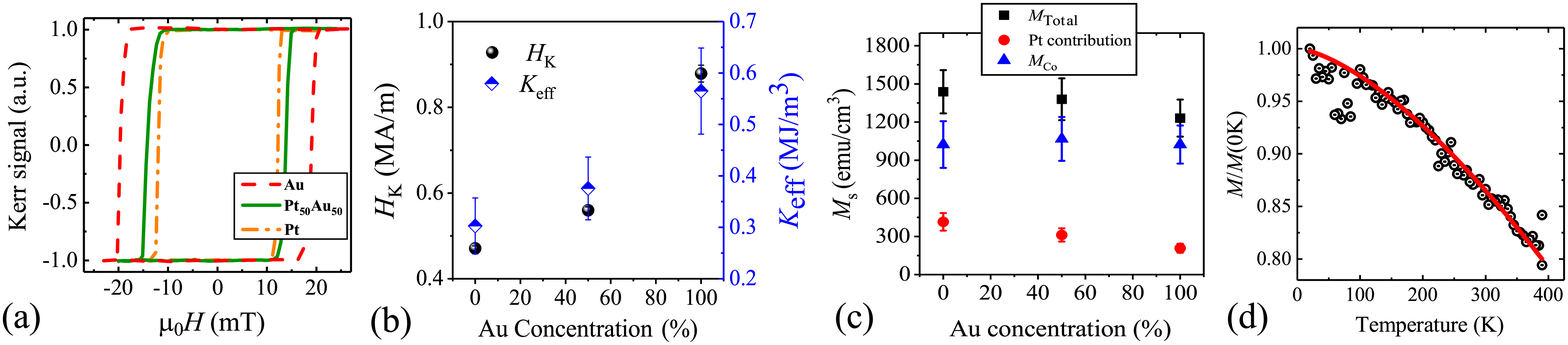}
  \caption{Magnetic characterization of Pt/Co/Au$_x$Pt$_{1-x}$ trilayers. (a) Polar MOKE loops. (b) Anisotropy field, $H_\mathrm{K}$, and effective anisotropy energy, $K_\mathrm{eff}$ both increase with Au concentration. (c) Saturation magnetization $M_\mathrm{s}$ supposing that all the magnetic moment is confined in the Co layer (black dots), extracted proximity-induced magnetism in Pt layers (red) and corrected Co layer values $M_\mathrm{Co}$ (blue). (d)  Variation of relative saturation magnetization with temperature for Pt/Co/Pt. The solid red line is a fit of the Bloch law.
  \label{magnetic}}
\end{figure*}

The quality of these epilayers was confirmed by x-ray diffraction (XRD) and transmission electron microscopy (TEM). The XRD spectra for $x = 0$, 0.5, and 1, shown in Fig.~\ref{structure}(a), are consistent with the previous study \cite{mihai}. In each case, a broad principal peak (due to the extreme thinness of the layers) is found close to $\theta = 40^\circ$, surrounded by satellite Pendell\"{o}sung fringes. This peak is very close to bulk Pt (111) Bragg peak at $\theta=39.74^\circ$, showing the matching of the multilayer structure with the substrate. The appearance of Pendell\"{o}sung fringes in the XRD of a multilayer emphasise the uniformity of the strain in the layers and smoothness of the interfaces \cite{thinfilmbook,tanner2013characterization}. The peak moves to slightly lower angles as Au is introduced into the upper layer, as might be expected from a Vegard law-type behaviour.

Cross-sections of the three samples are shown as dark field high-resolution transmission electron microscopy (HRTEM) images in Fig.~\ref{structure}(b). Lattice fringes are visible extending coherently across all three layers, emphasising the epitaxial nature of the samples. Energy dispersive x-ray analyses of each of the three samples are shown in Fig.~\ref{structure}(c), showing distinct layers. The lattice spacing in the vertical direction, $d_{111}$, can be evaluated using position of the main peak in the XRD pattern and Bragg's law, or by averaging the lattice spacings in the cross-sectional TEM images. Comparing evaluated values from both measurements, there is just about 2\% difference in $d_{111}$ between them (within error bars) which is a good agreement.

The contrast in TEM images is related to the atomic number inducing a darker appearance for the Co layer in comparison with Pt and Au. Investigating the changes of greyness in vertical direction of images results in Co thicknesses of 1.03~nm, 1.26~nm and 1.17~nm for $x=1$, 0.5 and 0, respectively. (The thickness was measured for stripes of $\sim0.4$~nm width across the entire image, and then the average was calculated). This is much higher than the nominal thickness of 0.6~nm. In fact, this is expected as TEM shows the additive effect of all atomic layers in the depth of FIB-cut lamella (which is usually $\sim50$~nm). So, when the electron beam goes through the TEM specimen, the thickness fluctuations of the thin layer will be added to the actual thickness resulting in the overestimation of thickness. It has also been reported previously than evaluating thickness from TEM cross-sections might result in overestimated thicknesses, sometimes even two times higher than the real value \cite{gabureac2014longrange}. Hence, x-ray reflectivity patterns of samples were fitted using the \textsc{GenX} package \cite{Bjoerck2007}, which yielded thicknesses very close to the nominal ones for the Co layers. Considering all these facts, the nominal thickness of 0.6~nm was taken to be a good estimate of the true thickness value for further analysis of parameters describing the samples.

All the films showed square magneto-optical Kerr effect (MOKE) hysteresis loops as a function of perpendicular magnetic field, confirming that the PMA generated at the Co-heavy metal interfaces is strong enough to overcome the shape anisotropy (Fig.~\ref{magnetic}(a)). Superconducting quantum interference device vibrating sample magnetometry (SQUID-VSM) was used to measure the anisotropy field $H_\mathrm{K}$, inferred from the in-plane saturation field, plotted in Fig.~\ref{magnetic}(b), and the saturation magnetization $M_\mathrm{s}$ shown in  Fig.~\ref{magnetic}(c). To calculate $M_\mathrm{s}$, it was assumed that all the measured moment is confined in the Co layer of thickness $t$. However, due to the proximity effects, induced magnetic moments also exist in the Pt layers. The induced moment in Pt can be estimated by assuming a negligible proximity effect in Au \cite{wilhelm2004magnetic}, and so the difference in moment between the $x=0$ and $x=1$ films corresponds to the induced moment in a single Pt layer. This is summarized in Fig.~\ref{magnetic}(c) which reveals the contribution of the magnetization in the Pt, to the constant magnetization of the Co layer ($1.0 \pm 0.1$~MA/m). This is comparable to the previously published values for Co/Pt multilayers with the same nominal thickness of cobalt \cite{lin1991magnetic, bandiera2012enhancement, lavrijsen2015asymmetric,metaxas, hrabec2014measuring, Miron_Nature2010}. On the other hand, if we consider thicknesses deduced from the TEM measurements, the saturation magnetization of mulilayers would be about 0.6-0.7~MA/m which is unusually low for supposedly 1~nm of Co \cite{lin1991magnetic,vanatka2015velocity}. This also emphasises the unreliability of thickness estimations from TEM cross-sections. Following this analysis, the effective anisotropy can be calculated by $K_\mathrm{eff} = \frac{1}{2} \mu_0 M_\mathrm{Co} H_\mathrm{K}$ that is also plotted in Fig.~\ref{magnetic}(b), which rises markedly with Au content $x$. The values, along with the other magnetic data for each sample, are given in Table~\ref{magproperties}. The temperature dependence of $M_\mathrm{Co}$ between 5 and 400~K for the $x = 0$ sample shown in Fig.~\ref{magnetic}(d) was fitted with the Bloch law\cite{chikazumibook1997},
\begin{equation}
\frac{M_{\mathrm{Co}(T)}}{M_0} = 1 - c \left( \frac{k_\mathrm{B} T}{J} \right)^{\frac{3}{2}},
\label{bloch}
\end{equation}
where $M_0$ is the magnetization at 0~K, $c = 0.0294$ for an fcc lattice, $k_\mathrm{B}$ is the Boltzmann constant and $J$ is the exchange integral. Knowing $J$, one can obtain the exchange stiffness as $A = Q J S^2 / a$, with $Q = 4$ as the number of atoms per unit cell for fcc, $S$ the spin quantum number, and $a = 0.355$~nm the lattice constant. The value we obtain is $A=17\pm1$~pJ/m. This reduction in $A$ with respect to the bulk value for Co of 30~pJ/m is in good agreement with our previous findings for Co layers of this thickness \cite{shepley2018}.

\begin{table*}[tb]
\caption{Magnetic properties of epitaxial Pt/Co/Au$_x$Pt$_{1-x}$ for the three different values of Au concentration $x$ studied here. These are saturation magnetization $M_\mathrm{CO}$, uniaxial anisotropy  constant $K_\mathrm{u} = K_\mathrm{eff} + \frac{1}{2} \mu_0 M_\mathrm{Co}^2$, DW width $\Delta = \sqrt{A/K_\mathrm{eff}}$ , Gilbert damping constant from creep and depinning fits $\alpha_\mathrm{exp}$, Walker breakdown field $\mu_0 H_\mathrm{W}$ calculated using DW mobility, Gilbert damping constant from micromagnetics $\alpha_\mathrm{\upmu M}$, depinning field $H_\mathrm{d}$ obtained from fitting with the creep law, thickness fluctuation $\delta $ used in the micromagnetics simulations, correlation length of the disorder $\xi$, strength of the pinning disorder $f_\mathrm{pin}/\xi$, and effective DMI constant $D_\mathrm{eff}$, as well as the inferred handedness of the chirality of the domain walls.} \label{magproperties}
\begin{ruledtabular}
   \begin{tabular}{c | c c c c c c c c c c c c c}
   $x$ & $M_\mathrm{Co}$ & $K_\mathrm{u}$ & $\Delta$ & $\alpha_\mathrm{exp}$ & $\mu_0 H_\mathrm{W}$ & $\alpha_\mathrm{\upmu M}$ & $\mu_0 H_\mathrm{d}$ & $\delta$ & $\xi$ & $f_\mathrm{pin}/\xi$ & $D_\mathrm{eff}$ & Chirality \\ \hline
    & MA/m & MJ/m$^3$ & nm & & mT & & mT & & nm & meV & mJ/m$^2$ &  \\ \hline
   1 & $1.0\pm0.1$ & $1.2\pm0.2$ & $5.5\pm0.9$ & $0.9\pm0.1$ & $280\pm10$ & $0.35\pm 0.05$& $72\pm2$ &$9\%$ & $25 \pm 3$ & $510 \pm 50$ & $-1.0\pm0.3$ & Left \\
   0.5 & $1.1\pm0.2$ & $1.1\pm0.2$  & $7\pm1$ & $0.4\pm0.1$ & $49\pm3$  & $0.35\pm 0.05$& $56\pm2$&$8.5\% $ & $21 \pm 2$ & $330 \pm 30$ & $-0.35\pm0.09$ & Left \\
   0 & $1.0\pm0.2$ & $1.0\pm0.2$ & $8\pm1$ & $0.2\pm0.1$ &  $11\pm1$ & $0.17\pm 0.01$& $39\pm2$&$ 7\%$ & $22 \pm 2$ & $240 \pm 20$ & $-0.07\pm0.07$ & None \\
   \end{tabular}
\end{ruledtabular}
\end{table*}

\section{Domain wall chirality}

Ferromagnet-heavy metal interfaces also exhibit interfacial DMI\cite{fert1991magnetic}, which enforces the N\'{e}el form of the DW profile with a fixed chirality\cite{Thiaville_DMI,benitez2015magnetic}. Here we have studied this property using the asymmetric bubble expansion method\cite{choe,hrabec2014measuring}. A bubble domain is nucleated and then expanded by an out-of-plane applied field $H_z$, whilst $H_\mathrm{DMI}$, the effective field induced at the DW by the DMI\cite{Thiaville_DMI} is enhanced or compensated by an in-plane applied field $H_x$, resulting in different creep-regime velocities for walls on either side of the bubble moving with or against the in-plane field \cite{Kabanov10IEEE,choe}. The effective micromagnetic DMI strength is given by $D_\mathrm{eff} = \mu_0 H_{\mathrm{DMI}} M_\mathrm{Co} \Delta$, where $\Delta$ is the DW width. To measure $H_\mathrm{DMI}$ we have used Kerr microscopy with a two-coil setup to apply $H_z$ and $H_x$ independently. We measure the DW creep velocity as a function of $H_x$ by pulsing $H_z$ with a fixed amplitude. The DW motion at low magnetic fields follows the creep law \cite{metaxas}, and its velocity can be expressed as
\begin{equation}\label{equ_velocity}
	v =v_0 \exp\left( \zeta H_z^{-\frac{1}{4}}\right)
\end{equation}
where $v_0$ is the velocity prefactor, $H_z$ is out-of-plane driving field, and $\zeta$ is the scaling coefficient
\begin{equation}\label{equ_zeta}
  \zeta=\zeta_0 \left[ \frac{\sigma(H_x)}{\sigma(0)} \right] ^{\frac{1}{4}}
\end{equation}
where $\zeta_0$ can be considered as a scaling constant for pinning properties independent of in-plane field. $\sigma$ is the N\'{e}el wall energy density \cite{choe,hrabec2014measuring}, which contains a term proportional to $|H_x + H_\mathrm{DMI}|$. Hence the highest wall energy, and (through Eqs~\ref{equ_velocity} and {\ref{equ_zeta}) minimum DW creep velocity $v$ corresponds to the point where $\mu_0H_x= -\mu_0H_{\mathrm{DMI}}$, i.e. the point at which external field transforms the N\'{e}el wall to a Bloch wall\cite{hrabec2014measuring}.

We performed bubble expansion measurements on all three epilayers, the results are shown in Fig.~\ref{dmi}. Measurements as a function of $H_z$ at given values of $H_x$ (not shown) yielded a $\ln v \propto H_z^{-1/4}$ scaling, confirming that measurements are in the creep regime. A typical Kerr difference micrograph of an expanded DW bubble in Pt/Co/Au film is displayed in inset of Fig.~\ref{dmi}(a), showing the asymmetric expansion. Examples of datasets for $v(H_x)$ for Pt, Pt$_{50}$Au$_{50}$, and Au capping layers are shown in Fig.~\ref{dmi}(b), (c), and (d), respectively. One can immediately see a shift of velocity minimum away from zero in-plane field, indicating the presence of a DMI. The solid lines correspond to the fits of Eq.~\ref{equ_velocity}. The effective DMI field for each sample, determined from these fits, is plotted in Fig.~\ref{dmi}(a) as a function of Au concentration $x$. It should be noted that the high uncertainty in the $D$ value for Au capped trilayer is due to not reaching the minimum velocity (\textit{i.e.} the inversion symmetry axis of the curve) cannot be seen, as a result of experimental limitations, and so our value is based just on the fit to the data that is shown. Negligible DMI in Pt/Co/Pt confirms that the system has an high degree of inversion symmetry for $x=0$, which is not a trivial result \cite{hrabec2014measuring,lavrijsen2015asymmetric}, since small differences in top and bottom interface structure can easily break inversion symmetry and lead to an appreciable DMI\cite{Wells2017}. The negligible DMI in this case can thus be expected to lead to walls of of Bloch form with no fixed chirality.

Also, Kim \textit{et al.} stated that this assumption that the velocity minimum occurs at $-\mu_0 H_\mathrm{DMI}$ only holds if the minimum velocity falls on the inversion symmetry axis of the velocity curve  \cite{Kim18NPGAsiaMat}. In other words, it holds if the asymmetric contribution to the velocity is negligible. In such a case, any contribution from chiral damping is negligible and DMI can be accounted for all the chiral behaviour in the sample. Examination of symmetric and asymmetric contributions of our velocity curves, showed that the latter is one order of magnitude smaller than the former, proving the dominance of the symmetric contribution and ruling out any chiral damping effect.

The effective DMI field and the strength of DMI constant $D_{\mathrm{eff}}$ increase with $x$. This is expected since the trilayers are becoming more inversion asymmetric. From the symmetries of the applied fields we deduce that for $x>0$, the DMI enforces left-handed chirality of the N\'{e}el walls, which is consistent with \textit{ab initio} calculations for Pt/Co interfaces \cite{yang2015anatomy}.

\begin{figure}[tb]
  \includegraphics[width=8.4cm]{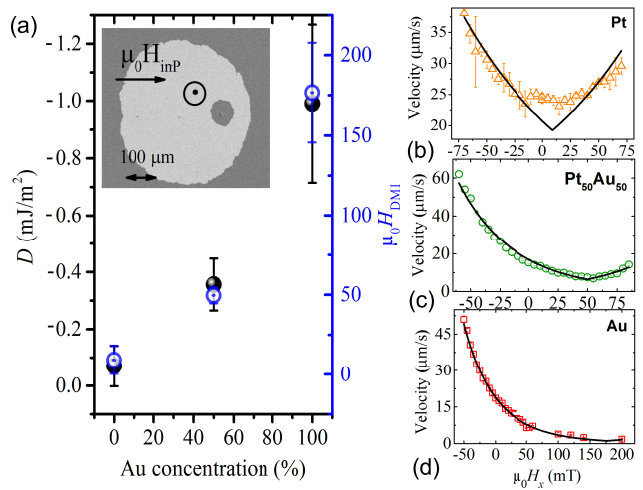}
  \caption{Asymmetric bubble expansion measurements. (a) Effective DMI field $\mu_0H_\mathrm{DMI}$ and effective DMI constant $D_\mathrm{eff}$ as a function of Au concentration $x$ in Pt/Co(6~\AA)/Au$_x$Pt$_{1-x}$ epilayers. An example of a Kerr difference image of the bubble expansion in Pt/Co/Au is shown as an inset. The values plotted were derived from the data in panels (b)-(d), which show DW velocity as a function of in-plane magnetic field $\mu_0H_x$ in case of Pt, Pt$_{50}$Au$_{50}$ and Au capping layers, respectively, for $\mu_0 H_z =7$, 8, and 10.5~mT.
  \label{dmi}}
\end{figure}

\section{Field-driven domain wall dynamics}

In the previous section, we considered some aspects of the creep regime, where the applied field is small with respect to the depinning field and DW motion is thermally activated. For applied fields exceeding the depinning field, DW dynamics first goes through a so-called depinning transition and then move in the viscous flow regime where temperature plays no role: the DW velocity is controlled by damping, parameterised by the Gilbert damping coefficient $\alpha$. We measured the DW velocity across both regimes by investigating bubble expansion driven by pulsed out-of-plane applied field using a fast micro-coil of diameter 1~mm that produced pulses with a 200~ns rise time. The DW velocity is then calculated by dividing the distance which the DW covered during the pulse by the pulse duration.
The velocities as a function of out-of-plane field $\mu_0H_z$ are shown in Fig.~\ref{dwmotion}(a). The DW velocity for a specific value of $H_z$ is rising with the broken inversion symmetry of the trilayer. The measured velocities cover the thermally activated creep regime, the depinning transition, and the beginning of the appearance of the expected linear change of velocity in the flow regime. The data were fitted simultaneously with the universal creep and depinning regime functions\cite{jeudy2017pinning}:
\begin{equation}\label{equ_creep}
 v(H) =
 \begin{cases}
   v(H_\mathrm{d})\exp\left(-\frac{\Delta E}{k_\mathrm{B}T}\right) & \textrm{Creep}\\
   \frac{v_\mathrm{T}(H_\mathrm{d})}{x_0}\left(\frac{H_z-H_\mathrm{d}}{H_\mathrm{d}}\right)^\beta & \textrm{Depinning}
 \end{cases}
\end{equation}
where $v(H_\mathrm{d})$ is the creep prefactor corresponding to the velocity at depinning, $T$ is the temperature, $k_\mathrm{B}$ is the Boltzmann constant, $H_\mathrm{d}$ is the depinning field, $\Delta E=k_\mathrm{B}T_\mathrm{d}[(H_z/H_\mathrm{d})^{-1/4}-1]$ is the pinning potential, $v_T=v(H_\mathrm{d})(T_\mathrm{d}/T)^{\psi}$ is the DW velocity in a disorder-free case, $x_0=0.65\pm0.04$ is a dimensionless constant\cite{DiazPardo17PRB}, and $\psi = 0.15$ and $\beta = 0.25$ are depinning exponents\cite{jeudy2017pinning}. Here we define the creep regime as $H \leq H_\mathrm{d}$ and the depinning regime as $H \gtrsim H_\mathrm{d}\left[ 1+(0.8(T_\mathrm{d}/T)^{-\psi})^{1/\beta}\right]$. The fitting results, which yield a good agreement with the acquired experimental data, are shown in Fig.~\ref{dwmotion}(a) and Fig.~\ref{dwmotion}(b). There is good agreement with both the creep law and with the depinning law for $H > H_\mathrm{d}$.

The depinning field $H_\mathrm{d}$ that separates the creep and depinning regimes of DW motion is marked on each curve in Fig.~\ref{dwmotion} by solid stars. The only-non-universal (\textit{i.e.} material and temperature dependent) parameters of the fits are the depinning field $H_\mathrm{d}$, depinning velocity $v_\mathrm{T}(H_\mathrm{d})$, and depinning temperature $T_\mathrm{d}$. All these three parameters are increasing with the Au concentration in the system (Fig.~\ref{dwmotion}(c-e)). Following Ref.~\onlinecite{jeudy2017pinning}, the material dependent parameters $H_\mathrm{d}$ and $T_\mathrm{d}$ can be used to estimate the variation with gold concentration of the coherence length ($\xi=\left[ (k_\mathrm{B} T_\mathrm{d})^2/(2 M_\mathrm{s} H_\mathrm{d} \sigma t^2) \right]^{1/3}$ ), and the pinning strength ($f_\mathrm{pin}\xi=\left[ (k_\mathrm{B} T_\mathrm{d})^3/(\sigma \xi t) \right]^{1/2}$). The results are reported in Table~\ref{magproperties}. It can be seen there that the correlation length of the disorder is almost constant($\xi = 23\pm 2$~nm), which suggests that the typical length scale of the variations of the pinning potential (i.e. of the fluctuation of DW energy) is independent of the gold concentration. Moreover, the increase of the pinning strength suggests that fluctuations of DW energy are larger at the CoAu interface than at the CoPt interface. This is compatible with a smearing of DW energy fluctuations at the CoPt interface due to intermixing, which do not exist at the CoAu interface since Co and Au do not form any alloy\cite{massalski1986binary}.

%The increase of pinning related parameters for larger $x$ could be understood by the fact that Co and Au do not form any alloy \cite{massalski1986binary} and that Au is only weakly magnetically polarizable \cite{wilhelm2004magnetic}, so the magnetic interface is much more abrupt than in the case of Co/Pt where the magnetic interface roughness is smeared out by the proximity effects.

\begin{figure}[tb]
  \includegraphics[width=8cm]{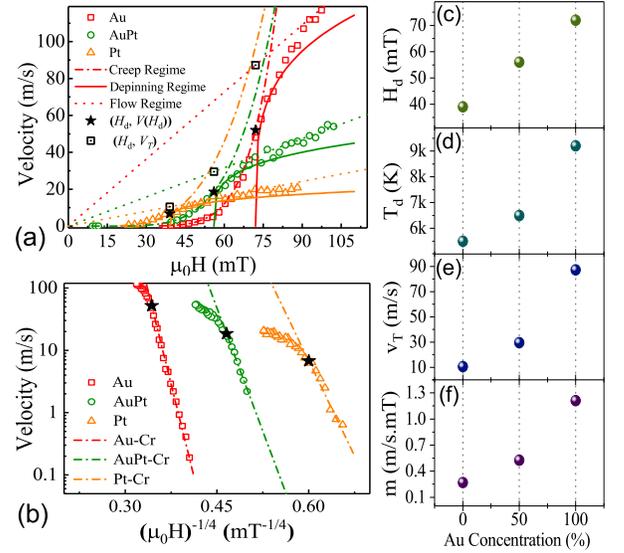}
  \caption{OoP-only field-driven DW motion for Pt/Co/Au$_x$Pt$_{1-x}$ epilayers. (a) Fits to the data by universal models for creep (dashed-dotted curve) and depinning regimes (solid curve) on a linear-linear plot. The dotted lines are linear fits to the viscous flow regime. (b) Semi-logarithmic plot of the same velocity data as a function of $(\mu_0H)^{-1/4}$ to show fitting to the creep regime more clearly. Data are shifted to the right respectively for clarity of presentation. In each case, the solid stars show the depinning field $H_\mathrm{d}$. (c)-(f) Changes of derived parameters with concentration of gold $x$: depinning field $H_\mathrm{d}$, depinning temperature $T_\mathrm{d}$, DW velocity in a film with pinning $v_T$, and DW mobility $m$, respectively.
  \label{dwmotion}}
\end{figure}

In each case the DW enters the flow regime, where $v \propto H_z$, at high fields. As stated before, $v_T$ is assumed to be the velocity of DW in the absence of pinning, hence DW mobility will be $m=\mu_0^{-1} v_T/H_\mathrm{d}$. Using this calculated $m$ [Fig.~\ref{dwmotion}(f)], a line can be drawn with this slope, going through origin, shown in Fig.~\ref{dwmotion}(a) as a dotted line. Non-trivially, the line falls onto the experimental data-points at high fields where $ v \propto H_z$. This is interesting, as the calculated DW mobility (that is a flow regime related parameter) is derived from creep and depinning fits to the experimental data which are measured with magnetic fields much lower than what viscous flow motion needs. In other words, we could learn about the flow regime and its parameters without a need to reach high, flow-related fields.
	
The standard one-dimensional (1D) model description of DW dynamics~\cite{Thiaville_DMI}, which describes the DW motion in terms of its position $q$ and internal spin configuration $\phi$, predicts that the DW mobility $m$ in the flow regime depends on Gilbert damping, $\alpha$, as follows:
\begin{equation}\label{equ_alpha}
  m =
  \begin{cases}
    \gamma\Delta/\alpha\qquad\qquad & \textrm{Steady flow} \\
    \gamma \Delta / \left( \alpha+\alpha^{-1} \right) & \textrm{Precessional flow}
  \end{cases}
\end{equation}
where $\gamma$ is the gyromagnetic ratio\cite{metaxas}. The Gilbert damping for the three layers, $\alpha_\mathrm{exp}$, determined from these dashed lines are given in Table~\ref{magproperties}. Eq.~\ref{equ_alpha}, which relates the DW mobility to $\alpha_\mathrm{exp}$, only gives a solution in the case of precesssional flow the trilayers capped with Pt and AuPt, whilst yielding a only steady flow solution for the epilayer where $x = 1$.

The Walker field, which separates the steady and precessional flow regimes, can be estimated by using relation \cite{Thiaville_DMI}
\begin{equation}H_\mathrm{W}=\alpha_\mathrm{exp}\sin\Phi_\mathrm{W}(H_\mathrm{DMI}-H_\mathrm{demag}\cos\Phi_\mathrm{W}),
\label{walker}
\end{equation}
where $H_\mathrm{demag}$ is the DW demagnetizing field, and $\cos\Phi_\mathrm{W} = ( [ H_\mathrm{DMI} / H_\mathrm{demag} ] - \sqrt{ [ H_\mathrm{DMI} / H_\mathrm{demag} ]^2 + 8} ) / 4.$ The values we obtain for $H_\mathrm{W}$ are also given in Table~\ref{magproperties}. The Walker field increases markedly with $x$, due to the stronger DMI as the inversion asymmetry is progressively broken. It can be seen that knowledge of the DMI field and Gilbert damping are needed to determine the regime of the DW dynamics with respect to the Walker field. Nevertheless, comparing these values with $H_\mathrm{d}$, we can see that our results are self-consistent, confirming which flow regime each sample enters on exceeding the depinning field.

\section{Micromagnetic simulation}

The Gilbert damping values $\alpha_\mathrm{exp}$ obtained from Eq.~\ref{equ_alpha} is unexpectedly large in the Pt/Co/Au sample \cite{metaxas}, suggesting that such a description, based on a 1D model, might not always be applicable in these systems. This could be due to a strong effect of disorder, since the 1D model usually assumes DW motion in perfectly homogeneous conditions.

\begin{figure}[tb]
  \includegraphics[width=7cm]{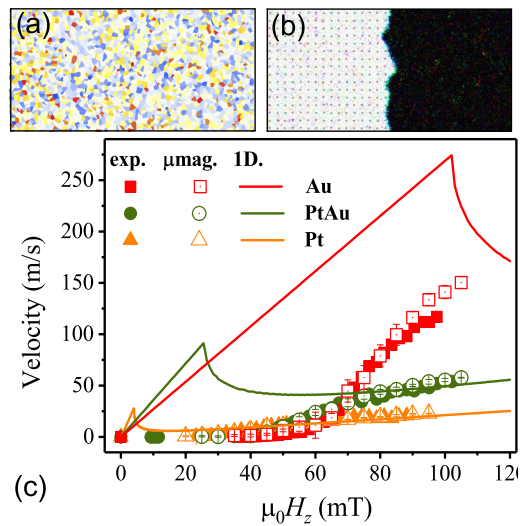}
  \caption{Field-induced DW motion in Pt/Co/Au$_x$Pt$_{1-x}$ epilayers. (a) and (b) Visualisations of the micromagnetic simulations for Pt/Co/Au with $\delta=8.5\%$, $\alpha=0.35$ at $t=0$ns. (c) Field-dependent velocity curves along with the simulation from micromagnetic simulations and 1D model calculations. It is shown that 1D model calculations cannot reproduce the data while the micromagnetic simulations can follow change in velocity precisely.
  \label{dwsim}}
\end{figure}

In order to test this conjecture, we performed numerical micromagnetic simulations using the \textsc{mumax}$^3$ package \cite{Vansteenkiste2014}. We used the experimentally measured magnetic parameters ($A$, $K_\mathrm{u}$, $M_\mathrm{Co}$, and $D_\mathrm{eff}$) and implemented disorder by means of Voronoi tessellation (see Fig.~\ref{dwsim}(a)) of the sort usually used to represent a grain structure \cite{Kim2017,Zeissler2017}. Since our trilayers are epitaxial they do not have a grain structure \textit{per se}, but there are magnetic layer thickness fluctuations, as revealed by the TEM cross-sections. In each region the micromagnetic parameters were changed as expected from a random thickness fluctuations that affect $K_\mathrm{u}$, $D_\mathrm{eff}$ and $M_\mathrm{Co}$\cite{Hrabec2017,Gross2018,Moretti2017}. The standard deviation of this fluctuation is given by the parameter $\delta$. The simulated sample is discretized in rectangular cells of $(2 \times 2 \times t)~\mathrm{nm}^3$, where $t$ is the thickness of the sample.  We consider a region of $1.0~\upmu \mathrm{m} \times 0.5~\upmu \mathrm{m}$, with periodic boundary condition along the $y$ direction. The DW is relaxed at the center as shown in Fig.~\ref{dwsim}(b). The `grain' size is set to 15~nm, which is a scale consistent with our TEM analysis here, and close to the values of $\xi$ given in Table~\ref{magproperties}. We simulate the DW dynamics for $50$ ns, in order to reach a stationary state in which the calculated DW velocity is not influenced by the simulation time. We performed different simulations combining different values of $\alpha$ and $\delta$. As all the experimental measurements were done at room temperature, the simulation temperature was set to $T=300$~K, accounted for by a thermal field of the form given by Brown \cite{Vansteenkiste2014,Brown1963}. The best fits are shown in Fig.~\ref{dwsim}(c) for the three samples. The description of the data with this approach is excellent. The corresponding best fitting parameters ($\alpha_\mathrm{\upmu M}$ and $\delta$) are reported in Table~\ref{magproperties}. The larger $\delta$ we had to introduce for the samples doped with Au reflects the higher values of $f_\mathrm{pin}$ due to the more abrupt interface between Co and Au, as previously commented on.

In every case, $\alpha_\mathrm{\upmu M} < \alpha_\mathrm{exp}$, although the largest difference is for $x = 1$. That damping parameter for Pt/Co/Au of is in reasonable agreement with the values reported for typical Pt/Co systems $\alpha\approx 0.3$\cite{metaxas}. For Pt/Co/PtAu and Pt/Co/Pt are slightly lower but just within an error bar of $\alpha_\mathrm{exp}$.  We calculated the prediction of the 1D model for $\alpha_\mathrm{\upmu M}$, also shown Fig.~\ref{dwsim}(c) as solid lines, which do not coincide with the data at all. The failure of the 1D model can be easily understood if we look at the depinning fields obtained from the creep fits. When $H_z \approx H_\mathrm{d}$ for $H_\mathrm{d} < H_\mathrm{W}$, as is the case for Pt/Co/Au, disorder has a strong effect on the DW dynamics, which can be reproduced successfully only with full micromagnetic simulations. On the other hand, if $H_z \gg H_\mathrm{d} > H_\mathrm{W}$ disorder has a smaller effect on the DW dynamics and the  predictions of the 1D model are in agreement with micromagnetic calculations in that field regime, as is the case for Pt/Co/Pt and Pt/Co/PtAu. Not taking into account the DW's internal dynamics (such as the presence and collision of Bloch lines) can only lead to an upper limit estimation for damping \cite{Moretti2017}, due to the fact that all dissipation mechanisms are--implicitly--attributed to damping in that case \cite{Yoshimura2015}.

The larger damping in the more strongly Au-doped samples could be due to a larger spin pumping~\cite{SpinPumping} at the Co/Au interface. An increased spin Hall effect was indeed recently observed in Pt/Au/Co stacks~\cite{Mann2017} and attributed to a larger spin transmission at the Au/Co interface, which could also affect the spin pumping mechanism. However, it is unlikely that it could explain alone the large difference with the Pt/Co/Pt sample.  Moreover, similar spin mixing conductances were observed in Pt/Co and Au/Co by FMR~\cite{Singh2017}. Non-local damping (dependent on the spin configuration) \cite{Hankiewicz2008,Zhang2009,weindler2014magnetic} could also play an important role for narrow DWs and could lead to an increased \textit{effective} Gilbert damping~\footnote{The inclusion of the non-local damping would imply a modification of the LLG equation~\cite{Hankiewicz2008,Zhang2009}. Thus the value fitted with the standard LLG must be seen as an effective damping, which averages out these inhomogeneous contributions.}. Hankiewicz \textit{et al.}, in particular, showed that non-local damping increases with the disorder scattering rate and is strongly enhanced by electron-electron interactions for weak spin polarization \cite{Hankiewicz2008}, as could occur at the Co/Au interface or in case of intermixing. Non-local effects will also be stronger for the larger magnetization gradients around the narrower walls found for trilayers containing more Au. Also, it was reported that for Au/Co interfaces the very interfacial atomic layers are playing the most important role in damping of the system, while for Pt/Co bilayers the effect can be extended to a few atomic layers \cite{barati2014}. So the observed higher thickness fluctuations (\textit{i.e.} interface roughness) and probable intermixing of layers for Pt/Co/Au trilayers of this study can affect the effective damping.

\section{Conclusions}

We have prepared and studied a series of Pt/Co/Pt$_{1-x}$Au$_x$ epitaxial trilayers with a well-defined crystallographic structure and controlled levels of proximity magnetism and inversion symmetry breaking. We have determined their static magnetic properties and the field-driven domain wall dynamics that they support. Our main findings are as follows. Comparing the measured moments of our samples allows us to estimate the degree of proximity magnetization in Pt, which is about one-third of the total moment in a Pt/Co(0.6~nm)/Pt trilayer. Adding Au increases the PMA, and also breaks the inversion symmetry and increases the overall DMI. This DMI leads to homochiral left-handed N\'{e}el domain walls, as predicted by theory for Pt/Co\cite{yang2015anatomy}, suggesting that the DMI arising from the Au interface is weak, mirroring findings for the spin-orbit torque \cite{ryu2014chiral}. Adding Au also increases the depinning field and temperature for these walls, which is consistent with a more abrupt magnetic interface in the absence of proximity magnetism. The field-driven domain wall dynamics, including the crossover from creep to viscous flow, can be described well by a micromagnetic model that takes appropriate account of the disorder in the sample whereas a simple 1-D model fails to do so. This micromagnetic modelling reveals lower values of Gilbert damping than the 1-D case, due to the proper treatment of pinning, although still rising up to $\alpha \approx 0.3$ as Au is added. The rise is likely to be due to higher interface transparency leading to increased spin pumping or to higher levels of non-local damping. The overall approach to the study of these trilayers used here could be also extended to studying the physics of skyrmions\cite{Fert2013,Nagaosa2013}, which contain the same physics as the chiral walls although expanded into two dimensions.

\begin{acknowledgments}
This work was supported by the EU via WALL network (grant number FP7-PEOPLE-2013-ITN 608031), the UK EPSRC (grant number EP/I011668/1), and the Agence Nationale de la Recherche (France) under Contract No. ANR-14-CE26-0012 (Ultrasky). The work by E. M. was supported by Projects MAT2014-52477-C5-4-P and MAT2017-87072- C4-1-P from the Spanish government, and Project No. SA090U16 from the Junta de Castilla y Leon.
\end{acknowledgments}

\bibliography{dwrreview_SM_KSR}

\end{document}